\documentclass[11pt]{article}
\usepackage[dvips]{graphicx}

\newtheorem{defn}{Definition}[section]
\newtheorem{thm}[defn]{Theorem}

\newcommand{\proof}{\textbf{Proof. }}
\newcommand{\qed}{\hbox{\rule{6pt}{6pt}} \vspace{1em}}

\begin{document}
\pagestyle{empty}
\vspace*{10mm}

\begin{center}
 {\Large \bf A quantum protocol to win the graph colouring game
  on all Hadamard graphs \\
}

\vspace{7mm}
 {\large \bf David Avis$^1$ } 
 {\large \bf Jun Hasegawa$^2$ }
 {\large \bf Yosuke Kikuchi$^3$ and }
 {\large \bf Yuuya Sasaki$^2$\\}
\vspace{2mm}
 {
   $^1$
  Department of Computer Science, McGill University,
  3480 University, Montreal, Quebec, Canada H3A 2A7,
  {\tt avis@cs.mcgill.ca}\\}
 {
   $^2$
   Department of Computer Science, Graduate School of
   Information Science and Technology,
   The University of Tokyo,
   7-3-1 Hongo, Bunkyo-ku, Tokyo 113-0033, Japan
  {\tt \{y\_sasaki, hasepyon\}@is.s.u-tokyo.ac.jp}\\}
 {
   $^3$
   ERATO QCI Project, 
   JST, 
   Hongo White Building, 
   Hongo 5-28-3, Bunkyo-ku, 
   Tokyo 113-0033, Japan, 
   {\tt kikuchi@qci.jst.go.jp}
   \\}
\end{center}

\noindent
{\bf Abstract : }
This paper deals with graph colouring games, an example of pseudo-telepathy,
in which
two provers can convince a verifier
that a graph $G$ is $c$-colourable 
where $c$ is less than the chromatic number of 
the graph. 
They win the game if they convince the verifier.
It is known that the players cannot win if they 
share only classical information,
but they can win in some cases by sharing entanglement. 
The smallest known graph where the players win in the quantum setting, but not in the classical setting, 
was found by Galliard, Tapp and Wolf and has 32,768 vertices.
It is a connected component of the Hadamard graph $G_N$ with $N=c=16$.
Their protocol applies only to Hadamard graphs where $N$ is a power of 2.
We propose a protocol that applies to all Hadamard graphs. Combined with a result of Frankl,
this shows that the players can win on any induced subgraph of $G_{12}$ having 1609 vertices, with $c=12$.
Combined with a result of Frankl and Rodl, our result shows that all sufficiently large Hadamard graphs
yield pseudo-telepathy games.

\section{Introduction}

  It is known that quantum entanglement allows for a phenomenon
  called pseudo-telepathy, that is, two parties pretend to be 
  endowed with telepathic powers, as described in a survey paper by
  Brassard, Broadbent and
  Tapp~\cite{brassard04}.
  For two parties $A$ and $B$, 
  a pair of question $(q_a, q_b)\in  Q_A \times Q_B$ are given
  and then a pair of answer $(a_a, a_b)\in A_A \times A_B$ are returned.
  In an initial phase, the parties can communicate with each other and
  share information.
  If the parties share an entangled state, then this setting is called
  shared entanglement.
    In the second phase, the parties are no longer allowed
  to communication with each other before revealing their answers.
  The parties {\em win} this instance,
  if  $a_a=a_b \Leftrightarrow q_a=q_a$.
  A protocol is successful with probability $p$ if it wins any instance that
  satisfies the promise with probability at least $p$.
  This exchange is called a {\em pseudo-telepathy game} if 
  there is a protocol that is successful with probability $1$ with 
  shared entanglement and 
  does not admit 
  such a protocol that is successful with probability $1$ without
  sharing entanglement.
   
  The graph colouring game is an example of 
  a pseudo-telepathy game (Brassard, Cleve and Tapp~\cite{brassard04},
Cleve, H{\o}yer, Toner and Watrous \cite{cleve04}).
  In a {\em graph colouring game}, 
there are two provers, called Alice and Bob, and a verifier.
A graph $G(V,E)$ and a integer $c$ is given to Alice and Bob.
Alice and Bob agree on a protocol to convince the verifier  
that $G$ is $c$-colourable.
The verifier sends $a\in V$ to Alice and sends $b\in V$ to Bob
such that $a=b$ or $(a,b)\in E$.
Alice and Bob are not permitted to communicate after receiving 
$a$ and $b$.
Alice sends the colour $c_A$ of $a$ to the verifier and 
Bob sends the colour $c_B$ of $b$ to the verifier.
Alice and Bob win if  
$a\neq b$ and $c_A\neq c_B$ or $a=b$ and $c_A=c_B$,
and lose otherwise.
They win the game if they convince the verifier.

The chromatic number $\chi(G)$ of a graph $G$ is the the smallest number of colours that can be
assigned to vertices such that no two adjacent vertices receive the same colour.
  If $c \geq \chi(G)$
  then there exists a protocol to win with probability $1$
  by using a colouring of $G$ with $c$ colours.
  Otherwise, Alice and Bob cannot win the game with probability $1$
  using classical methods \cite{brassard99}.
Using shared entanglement however, there are graphs where they can win in this situation.
The Hadamard graphs, defined by Ito
  ~\cite{ito85-1,ito85-2}, provide such examples.

  Let $N=4k$ for any positive integer $k$.
  The {\em Hadamard graph} 
  $G_N$ is defined as the graph whose vertex set $V_N=\{0,1\}^N$
  and edge set $E_N=\{(u,v)\in V_N^2|d_H(u,v)=N/2\}$,
  where $d_H(u,v)$ means Hamming distance of $u$ and $v$.
  Hadamard graphs are related to Hadamard matrices, which 
are an important object of study in 
  combinatorics, especially design theory (e.g., see Stinson~\cite{stinson}).
  One of the open problems is to decide whether for every $k$,
  there exists a Hadamard matrix of order $4k$. 
  
  With shared entanglement Brassard , Cleve and Tapp ~\cite{brassard99}
showed that Alice and Bob
  win the graph colouring game with probability $1$ for $G_{2^n}$ and $c=2^n$
  by using the Deutsch-Jozsa protocol 
  and by sharing an entangled state.
  A result of Frankl and Rodl ~\cite{fr87} (Theorem 1.11) implies that for all large enough $n$, $\chi(G_{2^n}) > 2^n$,
and so asymptotically the game is an example
of a pseudo-telepathy game.
Galliard, Tapp and Wolf ~\cite{galliard0211} showed that this was already the case for $n=16$
using a rather complicated combinatorial argument. 
  Thus the graph colouring game for $G_{16}$
  and $c=16$ is a pseudo-telepathy game.

We extend these results in this paper to all Hadamard graphs. 
In the next section we state known results on 
the chromatic number of these graphs. 
In Section 3, we design a protocol to win the graph colouring game for 
all Hadamard graph with probability $1$.
Combing these results 
it is shown that the graph colouring game for $G_{12}$ 
with $c=12$
  is a pseudo-telepathy game. Furthermore this is the smallest value of $N$
for which
  $G_{N}$
  is a pseudo-telepathy game with $c=N$, and this holds for any induced subgraph with at least
1069 vertices..
  The concluding section proposes 
  a definition of quantum chromatic number and gives some open problems.

\section{Chromatic number of Hadamard graphs} 
  It is easily seen that $\chi(G_4)=4$ 
  and Ito~\cite{ito85-2} proved $\chi(G_8)=8$.
Given a graph $G=(V,E)$, the {\it independence number} of the graph, denoted $\alpha(G)$, is the
cardinality of the largest subset of vertices such that no two of them are joined by
an edge. 
Let $p \geq 3$ be an odd prime, $q \ge 1$, and $k= p^q$.
Frankl~\cite{frankl84} showed that 
$$
\displaystyle{
\alpha(G_{4k}) =
4 \sum_{i=0}^{k-1} {{4k-1} \choose i } < \frac{4^{4k}}{3^{3k}}.
}
$$

When $p=3$ and $q=1$ we get $\alpha(G_{12})=268$.
An elementary result of graph theory is that $\chi(G) \geq |V| / \alpha(G) $.
Therefore  $\chi(G_{12}) \geq 4096/268 >12$.
In fact $G_N$ consists of two identical connected components, each having independence number
half of that of $G_N$. 
Let $H$ be any induced subgraph, having 1609 vertices,
of one of the connected components of $G_{12}$.
(In an induced subgraph, two vertices are adjacent if and only if they are adjacent in the original graph.)
Then $\chi(H) \geq 1609/134 > 12$.


Since $\chi(G_4)=4$, $G_4$ is not a pseudo-telepathy game with $c=4$.
Similarly,  $G_8$ is not a pseudo-telepathy game with $c=8$.
However we will see that $H$ is a pseudo-telepathy game with $c=12$.
 
\section{A protocol with probability $1$ using QFT}
In this section we extend the protocol 
by Brassard, Cleve and Tapp~\cite{brassard99}.
Their protocol employs the quantum Hadamard transform
while our protocol employs the quantum Fourier transform (QFT) with any order,
which can be exactly done as shown by Mosca and Zalka~\cite{mosca04}.

We describe a protocol such that Alice and Bob win the 
graph colouring game of $G_N$ with probability $1$
where $2^{n-1} < N \le 2^n$.
In this protocol, we use the following two operations
$\mbox{QFT}_N$ and $P_{l_i}$:
$\mathrm{QFT}_N$ is a \textit{general}
quantum Fourier transform with order $N$,
not necessarily $2^n$, defined as
\begin{equation}
 \mbox{QFT}_N:|i\rangle
  \longrightarrow 
  \displaystyle{
  \frac{1}{\sqrt{N}}
  \sum_{j=0}^{N-1}(\omega)^{i \cdot j }|j\rangle},
\end{equation}
where
$\omega = \exp \left(
		\frac{2\pi\sqrt{-1}}{N}
	       \right)$.  
Mosca and Zalka~\cite{mosca04} show that 
this QFT with any order can be performed exactly.
The operation $P_{l_i}$ is a phase shift
corresponding to the $i$-th bit of an input string $l$
\begin{equation}
 P_{l_i} : |i\rangle \mapsto (-1)^{l_i}|i\rangle.
 \end{equation}
Our protocol has four steps.
Alice and Bob can communicate with each other at only step 1.\\[0.5pc]
%

\noindent
{\bf Step 1: Prepare initial state $|\Psi_{AB}\rangle$}\\
In step I, Alice and Bob prepare $2n$-qubits $|0\rangle^{\otimes n}
\otimes |0\rangle^{\otimes n}$.
Alice has the first $n$-qubits and Bob has the second.
Alice first applies $\mathrm{QFT}_N$ to her $N$-qubits:
\begin{eqnarray}
 |0\rangle^{\otimes n} \otimes |0\rangle^{\otimes n}
  \stackrel{QFT_N}{\mapsto}
  \frac{1}{\sqrt{N}}\sum_{i=0}^{N-1} |i\rangle
  \otimes |0\rangle.
\end{eqnarray}
Bob then applies controlled-NOT operations to his $n$-qubits
with Alice's qubits as control qubits
for sharing the initial entanglement states:
\begin{eqnarray}
  \frac{1}{\sqrt{N}}\sum_{i=0}^{N-1}|i\rangle
  \otimes |i\rangle =: |\Psi_{AB}\rangle.
\end{eqnarray}



%

\noindent
{\bf Step 2: Apply phaseshift $P_{a_i}$ and $P_{b_i}$}\\
Alice and Bob compute
 \begin{equation}
      (P_{a_i}\otimes P_{b_i})|\Psi_{AB}\rangle
      =
      \displaystyle{
      \frac{1}{\sqrt{N}}
      \sum_{i=0}^{N-1}(-1)^{a_i\oplus b_i}|i\rangle \otimes |i\rangle.
      }
 \end{equation}

\noindent
{\bf Step 3: Apply the general quantum Fourier transform} \\
Alice and Bob compute  
    \begin{equation}
      \begin{array}{l}
       \left(\mbox{QFT}_N\otimes \mbox{QFT}_N^{-1}
       \right)
       \left(
                P_{a_i} \otimes P_{b_i}	\right)
	|\Psi_{AB}\rangle\\
        =
        \displaystyle{
	  \left(\frac{1}{\sqrt{N}}\right)^3
	        \sum_{j_A=0}^{N-1}
		\sum_{j_B=0}^{N-1}
		\sum_{i=0}^{N-1}(\omega)^{i\cdot (j_A-j_B)}
		      (-1)^{{a_i}\oplus {b_i}}
		      |j_A\rangle \otimes |j_B\rangle.
	}
	\end{array}
    \end{equation}

\noindent
{\bf Step 4: Measure}\\
 Alice and Bob measure  her(his) qubits
using the computational bases and
each obtain one of $N$ basis states.
They output colours corresponding to their measurement result
to verifier.
\\[1pc]

We have the following theorem for our protocol.
\begin{thm}\label{thm:winning}
     Alice and Bob win the game with probability $1$ by our protocol. 
\end{thm}
\proof
Suppose that Alice(Bob) receives $a(b)$ and sends
$c_A(c_B)$.
The probability that Alice and Bob obtain
basis states $|j\rangle \otimes |j\rangle$ after measurement is
   $$
      \begin{array}{rl}
       \displaystyle{\left|
	  \langle j| \otimes \langle j|
	\left(\mbox{QFT}_N\otimes \mbox{QFT}_N^{-1}
	\right)
	\left(
        P_{a_i} \otimes P_{b_i}
	\right)
	|\Psi_{AB}\rangle\right|^2}
	= \displaystyle{\left|\left(\frac{1}{\sqrt{N}}\right)^3
	                       \sum_{i=0}^{N-1}(-1)^{a_i\oplus b_i}\right|^2} .
	   \end{array}
    $$
In case of $a = b$,
it holds that $a_i\oplus b_i=0$ for any $i$.
Thus the probability of $c_A=c_B$ is
$$
       \mbox{Pr}[c_A=c_B]
			= 1.
$$
On the other case of $a \neq b$,
it holds that $d_H(a, b)=\frac{N}{2}$, because of the definition
of the Hadamard graph. It means that
    $$
       \mbox{Pr}[c_A=c_B]
			= 0.
    $$
\qed

By combining Frankl's result~\cite{frankl84} and Theorem~\ref{thm:winning},
there is a gap between the shared entanglement setting and otherwise for $G_{12}$, and 
for the smaller subgraph $H$ mentioned in the previous subsection.
We have obtained the following result. 
\begin{thm}\label{for-G_{12}}
The smallest Hadamard graph $G_N$ such that
the graph colouring game is a pseudo-telepathy game with $c=N$ is $G_{12}$.
Any of its induced subgraphs with 1609 vertices also has this property.
\end{thm}
Godsil and Newman have proved that a Hadamard graph $G_N$ 
has chromatic number strictly larger than $N$
whenever $N=4m > 8$~\cite{gosil05}.
Then next result holds.
\begin{thm}\label{for-G_4p^q}
     The graph colouring game for Hadamard graph $G_{4m}$
     is a pseudo-telepathy game with $c=4m$ for all $m\geq 3$.
\end{thm}

The final statement of this theorem uses the result of Frankl and Rodl ~\cite{fr87} mentioned in Section 1. 

\section{Concluding remarks} 
In this paper, we have dealt with two party case for the quantum colouring game.
It may be interesting to investigate the multi-party case for quantum colouring
game.

The chromatic number $\chi(G)$ of a graph $G$
 is equal to the minimum number of colours such that
 Alice and Bob win the graph colouring game for $G$ 
 with probability $1$
without shared entanglement.
Patrick Hayden [private communication] suggested
we define the quantum chromatic number $\chi_Q(G)$
as the minimum number of colours such that
 Alice and Bob win the graph colouring game for $G$, using shared entanglement, 
 with probability $1$.
 It is easy to see that $\chi_Q(G)\le \chi(G)$,
and the pseudo-telepathy graph colouring game is
concerned with graphs with $\chi_Q(G)<\chi(G)$.
Characterizing such graphs $G$ would be interesting from 
both the standpoint of quantum communication
and of combinatorics. What is the smallest such graph?

For Hadamard graphs $G_{4p^q}$, there is an exponential gap between 
the chromatic number and the quantum chromatic number.
What is the largest such gap
as a function of the number of vertices of $G$?

\section*{Acknowledgments}
The authors would like to thank  Anne Broadbent, 
Patrick Hayden, 
Hiroshi Imai and Francois Le Gall
for their helpful comments and discussions.

\end{document}